\documentclass[aps,twocolumn,showpacs,preprintnumbers,amsmath,amssymb,pre]{revtex4-1}
\usepackage{epsf,amsmath,amssymb,amsfonts,verbatim,color,multirow,pifont}
\usepackage{graphicx}

\begin{document}
\title{Stochastic Thermodynamics and Hierarchy of Fluctuation Theorems with Multiple Reservoirs}
\author{Jae Sung Lee} \email{jslee@kias.re.kr}
\author{Hyunggyu~Park} \email{hgpark@kias.re.kr}
\affiliation{{School of Physics and Quantum Universe Center, Korea Institute for
Advanced Study, Seoul 02455, Korea}}

\newcommand{\revise}[1]{{\color{red}#1}}

\date{\today}

\begin{abstract}
We reformulate stochastic thermodynamics in terms of noise realizations for Langevin systems in contact with multiple reservoirs and investigated the structure of the second laws of thermodynamics. We derive a hierarchy of fluctuation theorems when one degree of freedom of the system is affected by multiple reservoirs simultaneously, that is, when noise mixing occurs. These theorems and the associated second laws of thermodynamics put stricter bounds on the thermodynamics of Langevin systems. We apply our results to a stochastic machine in noise-mixing environments and demonstrate that our new bounds play a crucial role in determining the potential function and performance of the machine.
\end{abstract}

\pacs{Kewords: multiple reservoirs, noise mixing, fluctuation theorems}
%\pacs{05.70.-a, 05.40.-a, 05.70.Ln, 02.50.-r Kewords}

\maketitle

\section{Introduction}

In the past two decades, thermodynamics has been extended to studies of stochastic thermal processes observed at the microscopic scale~\cite{Verley, Esposito, Garcia, Evans, Gallavotti, Jarzynski, Kurchan, Lebowitz, Crooks}. These processes are referred to as stochastic thermodynamics (ST). The main achievement in ST has been the discovery of the generalized second laws of thermodynamics, which is formulated by fluctuation theorems~\cite{Jarzynski, Crooks}. As the original second law of thermodynamics initiated studies of macroscopic thermal machines, such as heat engines or refrigerators, ST studies focus on devising various thermal machines with multiple reservoirs at microscopic~\cite{Blickle, Martinez, Su, bacteria} and even atomic~\cite{Johannes} scales. Therefore, it is important to establish a thermodynamic formulation of a stochastic system in contact with multiple reservoirs via a consistent ST framework.

However, theoretical difficulties arise when one degree of freedom of the machine is affected by multiple reservoirs simultaneously, that is, when thermal-noise mixing occurs. This situation can be easily encountered experimentally, for example, when a Brownian particle is immersed in a liquid with a temperature gradient~\cite{Schmidt}. The noise-mixing setup has also been proposed for a nano-sized molecular motor~\cite{Broeck} and refrigerator~\cite{Kawai}, where the rotor effectively described by a single degree of freedom is simultaneously connected to two thermal reservoirs separated by a membrane. In addition, chemical reaction between many chemical species via multiple reaction channels can be formulated as noise-mixing chemical Langevin equation~\cite{Horowitz}.

% % % % % % % % % % % % % % % % % %
\begin{figure}
\centering
\includegraphics[width=0.99\linewidth]{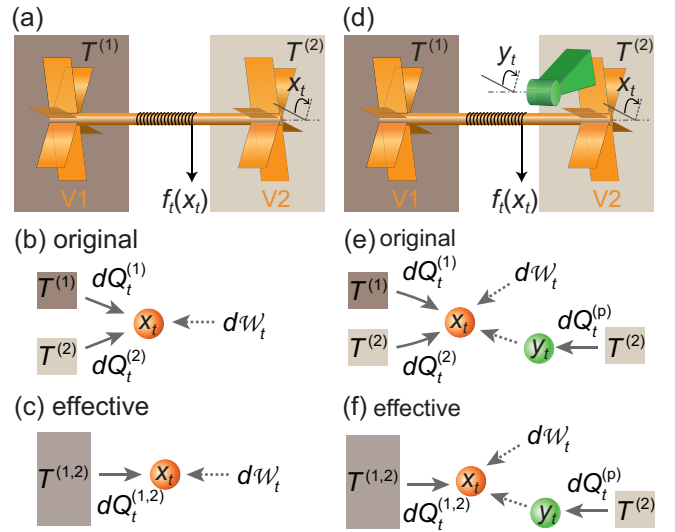}
\caption{ Steady-state heat machine in the noise-mixing environment. (a) Illustration of the vane system. Two symmetric vanes, V1 and V2, are connected by a rigid axle and immersed in heat reservoirs at temperatures of $T^{(1)}$ and $T^{(2)}$, respectively. $x_t$ is angle of the axle and $f_t(x_t)$ is an external force. (b, c) are simpler schematic diagrams of the vane system connected to two different reservoirs and one effective reservoir, respectively, where $dQ_{t}^{(1,2)}= dQ_{t}^{(1)}+ dQ_{t}^{(2)}$. (d) Illustration of the FSR-type model. (e, f) are the simpler schematic diagrams of the FSR-type model for the original and effective reservoirs, respectively. } \label{transducer}
\end{figure}
% % % % % % % % % % % % % % % % % % % % %

When noise mixing occurs, the conventional ST approach does not reveal all thermodynamic constraints enforced by the second laws of thermodynamics. For example, the total entropy production (EP) is underestimated~\cite{Broeck1}, and the simple overdamped limit is not applicable~\cite{Murashita}. There are two causes for this problem: i) the total EP for the noise-mixing system could not be obtained by the standard ST based on a probability ratio of system trajectories~\cite{Seifert2005, Schnakenberg, Hatano, Speck_Seifert, Spinney, Spinney1, P1,P2,Tome, Sagawa, Lee} and ii) the full structure of the second laws of thermodynamics for the noise-mixing system has not yet been investigated systematically.

In this study, we establish a ST formulation in terms of noise realizations instead of system trajectories. Our formulation applies to both mixing and non-mixing situations. In the mixing case, many noise realizations correspond to a single system trajectory due to multiple noises affecting the same degree of freedom of the system. This approach allows us to classify a single system trajectory into many {\em noise trajectories} with different noise realizations. We derive a hierarchy of fluctuation theorems based on these noise trajectories and the associated second laws of thermodynamics. We find that the total EP could be divided into two parts, the effective and mixing EP, each of which satisfies the fluctuation theorem along with the total EP. Therefore, thermal processes are more constrained by these additional second laws in the noise-mixing environment, causing a significant performance-bound reduction in a steady-state thermal machine.

\section{Reformulation of path probability in terms of noise realizations}

We first consider the simplest noise-mixing example, the vane system, shown in figure~\ref{transducer}(a), which appears in a prototypical Feynman-Smoluchowski ratchet (FSR)~\cite{Feynman, Smoluchowski}. One set of vanes, denoted by V1 (V2), is immersed in a heat reservoir $1$ ($2$) at a temperature of $T^{(1)}$ ($T^{(2)}$). V1 and V2 are connected by a rigid axle, such that the rotational motion of the vane system at time $t$ can be described by a single degree of freedom, the angle of the axle $x_t$. Because V1 and V2 are affected by random noise from the respective reservoirs, $x_t$ is a stochastic variable. $f_t(x_t)$ is an external (conservative/non-conservative/time-dependent) force applied to the vane system. Figure~\ref{transducer}(b) is a simpler schematic diagram of figure~\ref{transducer}(a); in this diagram, $dQ_{t}^{(\nu)}$ represents heat transferred from the reservoirs $\nu$ to the vane system, and $d{\cal W}_t$ is the work done by $f_t(x_t)$ during an infinitesimal time gap between $t$ and $t+dt$.

The Langevin equation for the vane system with unit mass (moment of inertia) can be written as ~\cite{MD, Derrida, Visco, Parrondo, Murashita}
\begin{eqnarray}
x_{t+dt} &=& x_t + v_t dt,\nonumber \\
v_{t+dt} &=& v_t + f_t(x_t)dt - \gamma^{(1,2)} v_t dt +dW_t^{(1,2)}, \label{eq:Langevin_eff}
\end{eqnarray}
where $v_t$ is the angular velocity at time $t$, and $\gamma^{(1,2)}=\gamma^{(1)}+\gamma^{(2)}$ is the composite damping coefficient, which is the sum of the damping coefficient $\gamma^{(\nu)}$ of each reservoir $\nu$. $dW_t^{(1,2)}=dW_t^{(1)}+dW_t^{(2)}$ is the composite Wiener process at $t$; it is the sum of two independent Wiener processes $dW_t^{(\nu)}$ of each reservoir $\nu$, satisfying $\langle dW_t^{(\nu)} \rangle =0$ and $\langle dW_t^{(\nu)}  dW_{t}^{(\nu^\prime)} \rangle =2D^{(\nu)} \delta_{\nu\nu^\prime} dt$ with $D^{(\nu)} =\gamma^{(\nu)}  T^{(\nu)}$, expressed using the Boltzmann unit ($k_B=1$).

We then define a state of the vane system at time $t$ as $q_t = (x_t, v_t)$ and consider its transition from $q_t$ to $q_{t+dt}$. The standard conditional probability, $\mathcal{P} (q_{t+dt}|q_t)$ for the time-forward transition of the system from $q_t$ to $q_{t+dt}$ is usually given by the Onsager-Machlup transition probability~\cite{Onsager-Machlup}. However, as the transition of our system is implemented by the two Wiener processes simultaneously, it is convenient to introduce new conditional probabilities subject to each noise realization $dW_t^{(1)}$ and $dW_t^{(2)}$ for a given $q_t$. Such a conditional probability should be given by the multiplication of the probabilities observing each Gaussian noise independently:
\begin{eqnarray}
\mathcal{P}(dW_t^{(1)},dW_t^{(2)}|q_t)=\prod_{\nu=1}^2 \frac{\mathbb{P}_t^{(\nu)}}{\sqrt{4D^{(\nu)}\pi dt}} \equiv \mathcal{P}_t^{(1),(2)} ,~~~~
\label{eq:new_path_prob}
\end{eqnarray}
where $\mathbb{P}_t^{(\nu)}=  \exp\left[-\left(dW_t^{(\nu)}  \right)^2/(4 D^{(\nu)} dt ) \right] $. Clearly, this probability does not depend on the initial state $q_t$ (no multiplicative noises); however, we keep this variable for later discussions. The final state $q_{t+dt}$ is determined by a pair of noises $(dW_t^{(1)}$, $dW_t^{(2)})$, starting from the initial state $q_t$, given by equation~\eqref{eq:Langevin_eff}. Note that there is an infinite number of pairs describing the same transition from $q_t$ to $q_{t+dt}$, with the sum $dW_t^{(1,2)}=dW_t^{(1)}$+ $dW_t^{(2)}$ invariant. In fact, this {\em degeneracy} allows us to derive additional fluctuation theorems.

As an irreversibility measure, EP involves the probability ratio between the time-forward and time-reverse trajectories~\cite{Seifert_review}. For the time-reverse process, we use $\widetilde{q}_{{\tilde t}}=(\widetilde{x}_{{\tilde t}}, \widetilde{v}_{{\tilde t}})$ for a state variable at time $\tilde t$, obeying dynamics identical to those of the time-forward process given in equation~\eqref{eq:Langevin_eff} with external force $\widetilde{f}_{\tilde t}=f_{t+dt}$ such that
\begin{eqnarray}
\widetilde{x}_{\tilde{t}+d\tilde{t}} &=& \widetilde{x}_{\tilde{t}} + \widetilde{v}_{\tilde{t}} d\tilde{t},\nonumber \\
\widetilde{v}_{\tilde{t}+d\tilde{t}} &=& \widetilde{v}_{\tilde{t}} + \widetilde{f}_{\tilde{t}}(\widetilde{x}_{\tilde{t}})d\tilde{t} - \gamma^{(1,2)}  \widetilde{v}_{\tilde{t}} d\tilde{t} + d\widetilde{W}_{\tilde{t}}^{(1,2)}, \label{eq:Langevin_eff_reverse}
\end{eqnarray}
where $d\widetilde{W}_{\tilde{t}}^{(1,2)}=d\widetilde{W}_{\tilde{t}}^{(1)}+d\widetilde{W}_{\tilde{t}}^{(2)}$ is the composite Wiener process for the time-reverse process at $\tilde{t}$, with the same statistics as
$\langle d\widetilde{W}_{\tilde{t}}^{(\nu)} \rangle =0$ and $\langle d\widetilde{W}_{\tilde{t}}^{(\nu)}
d\widetilde{W}_{\tilde{t}}^{(\nu^\prime)} \rangle =2D^{(\nu)} \delta_{\nu\nu^\prime} dt$ with $D^{(\nu)} =\gamma^{(\nu)}  T^{(\nu)}$.
 The conditional probability for the time-reverse transition from $\widetilde{q}_{{\tilde t}}$ to $\widetilde{q}_{{\tilde t}+d{\tilde t}}$ can be written as
\begin{eqnarray}
\widetilde{\mathcal{P}}(d\widetilde{W}_{\tilde t}^{(1)},d\widetilde{W}_{\tilde t}^{(2)}|\widetilde{q}_{\tilde t})=\prod_{\nu=1}^2
\frac{\widetilde{\mathbb{P}}_{\tilde t}^{(\nu)} }{\sqrt{4D^{(\nu)}\pi dt}} \equiv \mathcal{P}_{\tilde t}^{(\widetilde{1}),(\widetilde{2})},~~~~
 \label{eq:new_path_prob_reverse}
\end{eqnarray}
where
$\widetilde{\mathbb{P}}_{{\tilde t}}^{(\nu)}=\exp\left[-\left(d\widetilde{W}_{{\tilde t}}^{(\nu)}  \right)^2/(4 D^{(\nu)} dt )  \right] $.

By setting $\widetilde{q}_{{\tilde t}}=\varepsilon q_{t+dt}$ and $\widetilde{q}_{{\tilde t+d{\tilde t}}}=\varepsilon q_{t}$ ($\varepsilon$ is the parity operator as $\varepsilon q= (x,-v)$) with $d{\tilde t}=dt$, we identify the time-reverse trajectory corresponding to
the time-forward one.
However, this approach does not fix each value of Gaussian noise $(d\widetilde{W}_{\tilde t}^{(1)}, d\widetilde{W}_{\tilde t}^{(2)})$, except that their sum $ d\widetilde{W}_{\tilde t}^{(1,2)}=d\widetilde{W}_{\tilde t}^{(1)}+ d\widetilde{W}_{\tilde t}^{(2)}$ is fixed as
\begin{equation}
 d\widetilde{W}_{\tilde t}^{(1,2)} = d{W}_{t}^{(1,2)} -\gamma^{(1,2)} (v_t +v_{t+dt}) dt +{\cal{O}}((dt)^2)~.
\end{equation}
We consider the following specific pair of noises as
\begin{equation}
d\widetilde{W}_{\tilde t}^{(\nu)}=d{W}_{t}^{(\nu)} -\gamma^{(\nu)} (v_t +  v_{t+dt})dt ~~\quad (\nu=1,2).
\label{eq:symmetric_condition_W}
\end{equation}
This choice is special in that the heat transfer in the time-forward transition is exactly reversed in the time-reverse transition for each reservoir:
\begin{equation}
d\widetilde{Q}_{\tilde t}^{(\nu)}= - dQ_{t}^{(\nu)}, \label{eq:symmetric_condition}
\end{equation}
where $dQ_{t}^{(\nu)} = v_t \circ  (-\gamma^{(\nu)} v_t dt + dW_t^{(\nu)})$ and $d\widetilde{Q}_{\tilde t}^{(\nu)} = \widetilde{v}_{\tilde t} \circ  (-\gamma^{(\nu)} \widetilde{v}_{\tilde t} d{\tilde t} + d\widetilde{W}_{\tilde t}^{(\nu)})$ represent heat transferred from reservoir $\nu$ to the system during $dt$ in the time-forward and the time-reverse transitions, respectively, and $\circ$ denotes Stratonovich multiplication~\cite{Sekimoto, Noh}. If we consider Hamiltonian dynamics for the total system, including heat reservoirs, the time-reverse trajectory of the total system must obey the time-reversal symmetry of heat transfers from each reservoir, so the above choice may be the most appropriate as a definition of the time-reverse process.

With this special pair of time-reverse noises, we define an irreversibility measure by the logarithmic ratio of probabilities for time-forward and time-reverse noise trajectories as
\begin{eqnarray}
dR_{t} \equiv \ln \frac{p_t(q_t) \mathcal{P}_t^{(1),(2)}}{{\tilde p}_{\tilde t}({\tilde q}_{\tilde t}) \mathcal{P}_{\tilde t}^{(\widetilde{1}),(\widetilde{2})}} = dS_{t} - \frac{dQ_{t}^{(1)}}{T^\textrm{(1)}}-\frac{dQ_{t}^{(2)}}{T^\textrm{(2)}}
:=dS_t^\textrm{tot},~~~ \label{eq:new_path_prob_ratio}
\end{eqnarray}
where $p_t$ (${\tilde p}_{\tilde t}$) is the probability distribution function for the time-forward (time-reverse) transition at time $t$ ($\tilde t$). With the choice of ${\tilde p}_{\tilde t}({\tilde q}_{\tilde t})=p_{t+dt}(q_{t+dt})$, we obtain the Shannon entropy change of the system as $ dS_{t}\equiv \ln (p_t/p_{t+dt})$~\cite{Seifert2005}. In the derivation of equation~\eqref{eq:new_path_prob_ratio}, we used the relationship $ dQ_t^{(\nu)}= \frac{1}{2} v_t \circ [d\widetilde{W}_{\tilde t}^{(\nu)}+d{W}_{t}^{(\nu)}]$, which is easily obtained from equations~\eqref{eq:symmetric_condition_W} and \eqref{eq:symmetric_condition}. Our result indicates that $dR_{t}$ can be identified as the total EP, $dS_t^\textrm{tot}$ (i.e., the sum of the system and environmental EP).

\section{Hierarchy of fluctuation theorems}

Now, we show that $dS_t^\textrm{tot}$ satisfies the fluctuation theorem:
\begin{eqnarray}
\langle e^{-dS_t^\textrm{tot}}  \rangle
&=& \int dq_t  \prod_{\nu=1}^2 d(dW_{t}^{(\nu)})~   p_t (q_t)\mathcal{P}_t^{(1),(2)} e^{- dS_t^\textrm{tot} } \nonumber \\
&=& \int d{\tilde q}_{\tilde t}   \prod_{\nu=1}^{2} d(d\widetilde{W}_{\tilde t}^{(\nu)})~  {\tilde p}_{\tilde t}({\tilde q}_{\tilde t}) \mathcal{P}_{\tilde t}^{(\widetilde{1}),(\widetilde{2})} =1, ~~~~\label{eq:FT_tot}
\end{eqnarray}
where the Jacobian for integral variable transformations can be easily shown to be unity~\cite{jacob} and the last equality is derived from the probability normalization. This fluctuation theorem holds for any initial condition, any duration, and any external force. Furthermore, it is straightforward to generalize it to the case of many degrees of freedom, where different or the same multiple reservoirs are associated with each degree of freedom. We note that Murashita and Esposito~\cite{Murashita} and Forgedby and Imparato~\cite{Fogedby} also presented the fluctuation theorem for the total EP. However, as the total EP was not defined by the path probabilities in their study, their derivation is considerably more complex than ours.

We can find another quantity satisfying the fluctuation theorem in this noise-mixing situation. We consider the conditional probability to find $dW_t^{(1,2)}$ regardless of each value of noise $dW_t^{(\nu)}$ for a given $q_t$, which can be obtained as
\begin{eqnarray}
\mathcal{P} (dW_t^{(1,2)}|q_t) &=&\int_{-\infty}^\infty  \prod_{\nu=1}^2 d(dW_{t}^{(\nu)})~  \delta_W \mathcal{P}_t^{(1),(2)} \nonumber \\
&=&\frac{1}{\sqrt{4D^{(1,2)}\pi dt}}  e^{-\frac{\left(dW_t^{(1,2)}\right)^2}{4D^{(1,2)}dt}  } \equiv \mathcal{P}_t^{(1,2)}, \label{eq:old_new_relation}
\end{eqnarray}
where $\delta_W \equiv \delta \left(dW_t^{(1,2)}-dW_t^{(1)}-dW_t^{(2)} \right)$ is the Dirac delta function and $D^{(1,2)}=D^{(1)}+D^{(2)}$. Note that the usual Onsager-Machlup transition probability $\mathcal{P} (q_{t+dt}|q_t)$~\cite{Onsager-Machlup} is identical to $\mathcal{P} (dW_t^{(1,2)}|q_t)$ with a proper Jacobian. Similarly, we obtain
\begin{equation}
\widetilde{\mathcal{P}} (d\widetilde{W}_{\tilde t}^{(1,2)}|\widetilde{q}_{\tilde t})
=\frac{1}{\sqrt{4D^{(1,2)}\pi dt}} e^{-\frac{\left(d\widetilde{W}_{\tilde t}^{(1,2)}\right)^2}{4D^{(1,2)}dt}} \equiv \mathcal{P}_{\tilde t}^{(\widetilde{1},\widetilde{2})}. \label{eq:old_new_relation_reverse}
\end{equation}
Then, the logarithmic ratio of probabilities of the two trajectories turns out to be the following:
\begin{equation}
 dR_{t}^\prime \equiv \ln \frac{p_t (q_t) \mathcal{P}_t^{(1,2)}}{{\tilde p}_{\tilde t}({\tilde q}_{\tilde t}) \mathcal{P}_{\tilde t}^{(\widetilde{1},\widetilde{2})}} = dS_{t} - \frac{dQ_{t}^{(1,2)}}{T^\textrm{(1,2)}}:=dS_t^\textrm{eff}, ~~~~~\label{eq:old_path_prob_ratio}
\end{equation}
where $dQ_{t}^{(1,2)} = dQ_{t}^{(1)}+dQ_{t}^{(2)}$ and $T^{(1,2)}=D^{(1,2)}/\gamma^{(1,2)}$. Equation~\eqref{eq:old_path_prob_ratio} describes the total EP of the system connected to a single heat reservoir with an effective temperature of $T^{\textrm{(1,2)}}$, as illustrated in figure~\ref{transducer}(c). Note that $T^{(1,2)}$ is always between $T^{(1)}$ and $T^{(2)}$.

We call $dR_{t}^\prime$ the effective EP, $dS_t^\textrm{eff}$. We are then able to prove another fluctuation theorem:
\begin{eqnarray}
\langle  e^{-dS_t^\textrm{eff}} \rangle
&=& \int dq_t \prod_{\nu=1}^2 d(dW_{t}^{(\nu)})~ p_t(q_t) \mathcal{P}_t^{(1),(2)} e^{-dS_t^\textrm{eff}} \nonumber \\
&=& \int dq_t d(dW_t^{(1,2)})~ p_t(q_t) \mathcal{P}_t^{(1,2)} e^{-dS_t^\textrm{eff}} \nonumber \\
&=& \int d{\tilde q}_{\tilde t}   d(d\widetilde{W}_{\tilde t}^{(1,2)})~  {\tilde p}_{\tilde t}({\tilde q}_{\tilde t})
 \mathcal{P}_{\tilde t}^{(\widetilde{1},\widetilde{2})}=1,
\label{eq:FT_eff}
\end{eqnarray}
where the second equality was derived from equation~\eqref{eq:old_new_relation} by integrating over $dW_{t}^{(\nu)}$ after inserting the identity of $\int d(dW_t^{(1,2)}) \delta_W$.

Most interesting is the third quantity, satisfying another fluctuation theorem. We call it the mixing EP and define it as
\begin{equation}
dS_t^\textrm{mix} := dS_t^\textrm{tot} - dS_t^\textrm{eff}. \label{eq:mixingEP}
\end{equation}
The mixing EP measures the extent of thermal noise mixing and is independent of the system EP, $dS_{t}$; it satisfies the following:
\begin{eqnarray}
\langle e^{-d S^\textrm{mix}_t} \rangle &=& \int dq_t \prod_{\nu=1}^2 d(dW^{(\nu)}) p_t (q_t)
\mathcal{P}_t^{(1),(2)}  e^{- dS_t^\textrm{tot} +dS_t^\textrm{eff}} \nonumber \\
&=& \int d{\tilde q}_{\tilde t}   \prod_{\nu=1}^{2} d(d\widetilde{W}_{\tilde t}^{(\nu)})~
{\tilde p}_{\tilde t}({\tilde q}_{\tilde t}) \mathcal{P}_{\tilde t}^{(\widetilde{1}),(\widetilde{2})} e^{dS_t^\textrm{eff}}\nonumber\\
&=&\int d{\tilde q}_{\tilde t}   d(d\widetilde{W}_{\tilde t}^{(1,2)})~  {\tilde p}_{\tilde t}({\tilde q}_{\tilde t})
 \mathcal{P}_{\tilde t}^{(\widetilde{1},\widetilde{2})} e^{dS_t^\textrm{eff}}\nonumber\\
&=&\int dq_t d(dW_t^{(1,2)})~ p_t(q_t) \mathcal{P}_t^{(1,2)} = 1.
~~~~~ \label{eq:FT_mix}
\end{eqnarray}

\begin{figure*}
\centering
\includegraphics[width=0.96\textwidth]{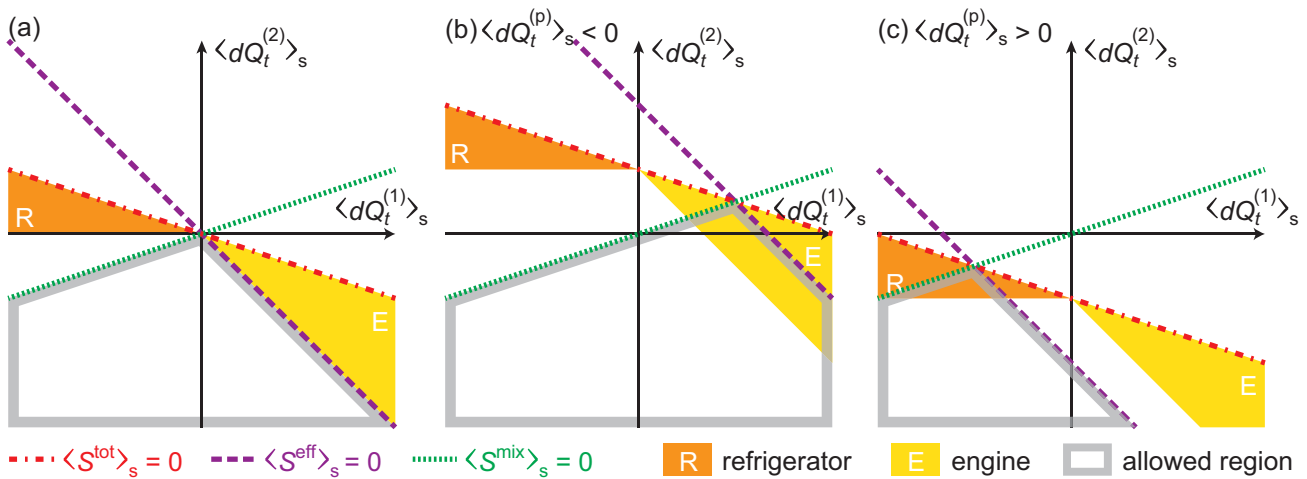}
\caption{ Three second laws of thermodynamics and the thermodynamically allowed region. (a) The vane system case. (b, c) FSR-type system with negative and positive $\langle dQ_t^{(\textrm{p})} \rangle_\textrm{s}$, respectively.} \label{fig:region}
\end{figure*}

In summary, we derived three integral fluctuation theorems for noise-mixing systems, which can be rephrased as follows. The total EP can be divided into the two parts, $dS_t^\textrm{mix}$ and $dS_t^\textrm{eff}$, each of which satisfies the fluctuation theorem as well as the total EP. This property is quite similar to the division of the total EP into the adiabatic and non-adiabatic EP in overdamped thermal systems with a single reservoir~\cite{Hatano,Speck_Seifert,Broeck2,Broeck3}. As $dS_t^\textrm{tot}$ and $dS_t^\textrm{eff}$ are written as the logarithmic ratio of two normalized probabilities, as in equations~\eqref{eq:new_path_prob_ratio} and \eqref{eq:old_path_prob_ratio} with the {\em involution} property~\cite{Broeck2}, it is trivial to derive the corresponding detailed fluctuation theorems, such as $P(dS_t^\textrm{tot})/P(-dS_t^\textrm{tot})=e^{dS_t^\textrm{tot}}$ and $P(dS_t^\textrm{eff})/P(-dS_t^\textrm{eff})=e^{dS_t^\textrm{eff}}$ in the steady state. However, the detailed fluctuation theorem does not hold for $dS_t^\textrm{mix}$ due to the lack of the involution property.

The three integral fluctuation theorems in equations~\eqref{eq:FT_tot}, \eqref{eq:FT_eff}, and \eqref{eq:FT_mix}, guarantee the three thermodynamic {\em second laws}, using Jensen's inequality~\cite{log_sum}:
\begin{eqnarray}
\langle dS_t^\textrm{tot}\rangle\geq 0,~~ \langle dS_t^\textrm{eff}\rangle\geq 0,~~ \langle dS_t^\textrm{mix}\rangle\geq 0,\label{eq:second_laws}
\end{eqnarray}
which can be summarized as follows:
\begin{equation}
\langle dS_t^\textrm{tot}\rangle\geq \langle dS_t^\textrm{eff}\rangle\geq 0. \label{eq:second_laws1}
\end{equation}
The non-negativity of $\langle dS_t^\textrm{mix} \rangle$ is interpreted as information loss in the heat reservoir coarse-graining procedure. Numerical test for equation~\eqref{eq:second_laws1} is presented in appendix A.
Note that $\langle dS_t^\textrm{mix}\rangle\geq 0$ can also be derived using the log sum inequality~\cite{log_sum}, because the trajectory probability $p_t (q_t) \mathcal{P}_t^{(1,2)}$ in the effective single-noise description is simply the log sum of the trajectory probabilities $p_t (q_t) \mathcal{P}_t^{(1),(2)}$ in the two-noise description. This property has been reported previously for the master equation system~\cite{Broeck2}. We also note that a richer hierarchy can be found between the total, effective, and mixing EPs in a general $n$-reservoir system $(n \ge 3)$ (see appendix B).

\section{Reduction of performance bound}

Each second law in equation~\eqref{eq:second_laws} constrains a thermodynamic process in a different manner. As expected, the effective and mixing EPs enforce a tighter performance bound for a noise-mixing machine than the total EP. As an example, consider the vane system illustrated in figure~\ref{transducer}(a) with $T^{(1)}>T^{(1,2)}>T^{(2)}$. The three second laws of thermodynamics in the steady state (or cyclic state) are given as follows:
\begin{eqnarray}
\langle dS_t^\textrm{tot} \rangle_\textrm{s}  &= &
-\frac{\langle dQ_t^{(1)} \rangle_\textrm{s}}{T^{(1)}} -\frac{\langle dQ_t^{(2)}\rangle_\textrm{s}}{T^{(2)}} \geq 0,  \label{eq:1EP}\\
\langle dS_t^\textrm{eff} \rangle_\textrm{s} &= &
-\frac{\langle dQ_t^{(1)} \rangle_\textrm{s} +\langle dQ_t^{(2)} \rangle_\textrm{s}}{{T^{(1,2)}}} \geq 0, \label{eq:2EP}
\\
\langle dS_t^\textrm{mix} \rangle_\textrm{s}  &= &
\frac{\langle dQ_t^{(1)} \rangle_\textrm{s}}{T^\textrm{(m,1)}} -\frac{\langle dQ_t^{(2)} \rangle_\textrm{s}}{T^\textrm{(m,2)}} \geq 0, \label{eq:3EP}
\end{eqnarray}
where $1/T^\textrm{(m,1)}=-1/T^\textrm{(1)}+1/T^\textrm{(1,2)}>0$ and $1/T^\textrm{(m,2)}=1/T^\textrm{(2)}-1/T^\textrm{(1,2)}>0$. Figure \ref{fig:region}(a) shows a generic thermodynamically allowed region for $\langle dQ_t^{(1)} \rangle_\textrm{s}$ and $\langle dQ_t^{(2)} \rangle_\textrm{s}$.
%Note that the allowed region is bounded by two second laws for the effective and mixed EPs.

From figure~\ref{fig:region}(a), we can deduce a general feature of the performance of the vane system. Defining the work extraction against the external force during the forward transition as $d{\cal W}_t$, energy conservation yields $d{\cal W}_t=dQ_t^{(1)}+dQ_t^{(2)}$. For a useful heat engine, we require that $\langle d{\cal W}_t\rangle_\textrm{s} =\langle dQ_t^{(1)}\rangle_\textrm{s} +\langle dQ_t^{(2)}\rangle_\textrm{s} > 0$. The second law of the total EP alone, equation~\eqref{eq:1EP}, permits a useful engine in the region shown in figure~\ref{fig:region}(a). However, this result is completely prohibited by the second law of the effective EP, equation~\eqref{eq:2EP}, manifesting the thermodynamic importance of additional second laws. In fact, our rigorous result extends the thermodynamic statement ``Work cannot be extracted from an engine connected to a single reservoir" to {the case of an engine connected {\em simultaneously} to multiple reservoirs which can be described by the Langevin equation~\eqref{eq:Langevin_eff}} without any spatial asymmetry~\cite{exp}. We also consider a refrigerator extracting heat from a low-temperature reservoir by external force, implying $\langle dQ_t^{(2)}\rangle_\textrm{s} >0$. This case does not contradict $\langle dS_t^\textrm{tot} \rangle_\textrm{s}$ alone, but is forbidden by the second law of the mixing EP, equation~\eqref{eq:3EP}.

It might be useful to add another system to interact with the vane system. One of the simplest examples is an FSR-type machine, as illustrated in figure~\ref{transducer}(d)~\cite{LeeJS1,LeeJS}. The added system (a pawl) is described by another stochastic variable, $y_t$, which is in contact with the reservoir $2$. It is straightforward to show the three fluctuation theorems similarly and the corresponding second laws in the steady state as follows:
\begin{eqnarray}
\langle dS_t^\textrm{tot} \rangle_\textrm{s}  &= &
-\frac{\langle dQ_t^{(1)} \rangle_\textrm{s}}{T^{(1)}} -\frac{\langle dQ_t^{(2)}\rangle_\textrm{s}}{T^{(2)}}
-\frac{\langle dQ_t^{(\textrm{p})}\rangle_\textrm{s}}{T^{(2)}}\geq 0,  \label{eq:1EPp}\\
\langle dS_t^\textrm{eff} \rangle_\textrm{s} &= &
-\frac{\langle dQ_t^{(1)} \rangle_\textrm{s} +\langle dQ_t^{(2)} \rangle_\textrm{s} }{{T^{(1,2)}}}
-\frac{\langle dQ_t^{(\textrm{p})}\rangle_\textrm{s}}{T^{(2)}}\geq 0, \label{eq:2EPp}
\\
\langle dS_t^\textrm{mix} \rangle_\textrm{s}  &= &
\frac{\langle dQ_t^{(1)} \rangle_\textrm{s}}{T^\textrm{(m,1)}} -\frac{\langle dQ_t^{(2)} \rangle_\textrm{s}}{T^\textrm{(m,2)}} \geq 0, \label{eq:3EPp}
\end{eqnarray}
where $dQ_t^{(\textrm{p})}$ is heat transferred from the reservoir $2$ to the pawl. Note that $dS_t^\textrm{mix}$ is independent of $dQ_t^\textrm{(p)}$, as the pawl dynamics are not related to noise mixing.

The function of this machine is determined by the sign of $\langle dQ_t^{(\textrm{p})} \rangle_\textrm{s}$. Figures~\ref{fig:region} (b) and (c) show the thermodynamically allowed region bounded by the three second laws for the negative and positive $\langle dQ_t^{(\textrm{p})} \rangle_\textrm{s}$, respectively. We require $\langle d{\cal W}_t\rangle_\textrm{s} =\langle dQ_t^{(1)}\rangle_\textrm{s} +\langle dQ_t^{(2)}\rangle_\textrm{s} +\langle dQ_t^{(\textrm{p})} \rangle_\textrm{s} > 0$ for a useful engine and $\langle dQ_t^{(2)}\rangle_\textrm{s}+\langle dQ_t^{(\textrm{p})} \rangle_\textrm{s}>0$ for a useful refrigerator. We demonstrate a thermodynamically allowed region for a useful engine in figure~\ref{fig:region} (b) and for a useful refrigerator in figure~\ref{fig:region} (c).

As described above, the thermodynamically allowed region is significantly reduced in the mixing system. However, we find that heat engine efficiency, defined as $\eta=\langle d{\cal W}_t\rangle_\textrm{s}/\langle dQ_t^{(1)}\rangle_\textrm{s}$, can still attain the Carnot efficiency $\eta_C=1-T^{(2)}/T^{(1)}$ at the uppermost corner of the allowed region in figure~\ref{fig:region} (b), where all three EPs vanish simultaneously. This efficiency is higher than the maximum efficiency of the effective system described by figure~\ref{transducer} (f), i.e., $\eta_C>1-T^{(2)}/T^{(1,2)}$, because $\langle dQ_t^{(2)}\rangle_\textrm{s}>0$ near the corner point of the noise-mixing system, which implies that the engine absorbs extra heat from reservoir 2 to convert into work. This problem is quite similar to the famous cooling-by-heating problem, in that it introduces an additional high-temperature reservoir~\cite{cooling,cooling1,cooling2}. The refrigerator in figure~\ref{fig:region} (c) also performs most efficiently at the uppermost corner, where the coefficient of performance, defined as $K=[\langle dQ_t^{(2)}\rangle_\textrm{s}+\langle dQ_t^{(\textrm{p})} \rangle_\textrm{s}]/\langle -d{\cal W}_t\rangle_\textrm{s}$ reaches a maximum of $K_{m}=1/[T^{(1)}/T^{(2)}-1]$. In this case, $\langle dQ_t^{(2)}\rangle_\textrm{s}$ is always negative, such that extra heat flows into reservoir 2 through the machine; thus, the refrigerator performs poorly compared to the effective system.

\section{Summary}

By reformulating the stochastic thermodynamics in terms of noise realizations, we derived three fluctuation theorems, which constrain a thermal process with multiple reservoirs more strictly using the effective and mixing EPs than it would using only the total EP. Nevertheless, attaining ideal efficiency remains possible. Further study of the possible setup for obtaining the Carnot efficiency is necessary.

\begin{acknowledgments}
This research was supported by the NRF grant No. 2017R1D1A1B06035497
(HP).
\end{acknowledgments}

\appendix

\renewcommand{\thefigure}{A\arabic{figure}}
\setcounter{figure}{0}

\section{Numerical test for equation~\eqref{eq:second_laws1}}

\begin{figure}
\centering
\includegraphics[width=0.99\linewidth]{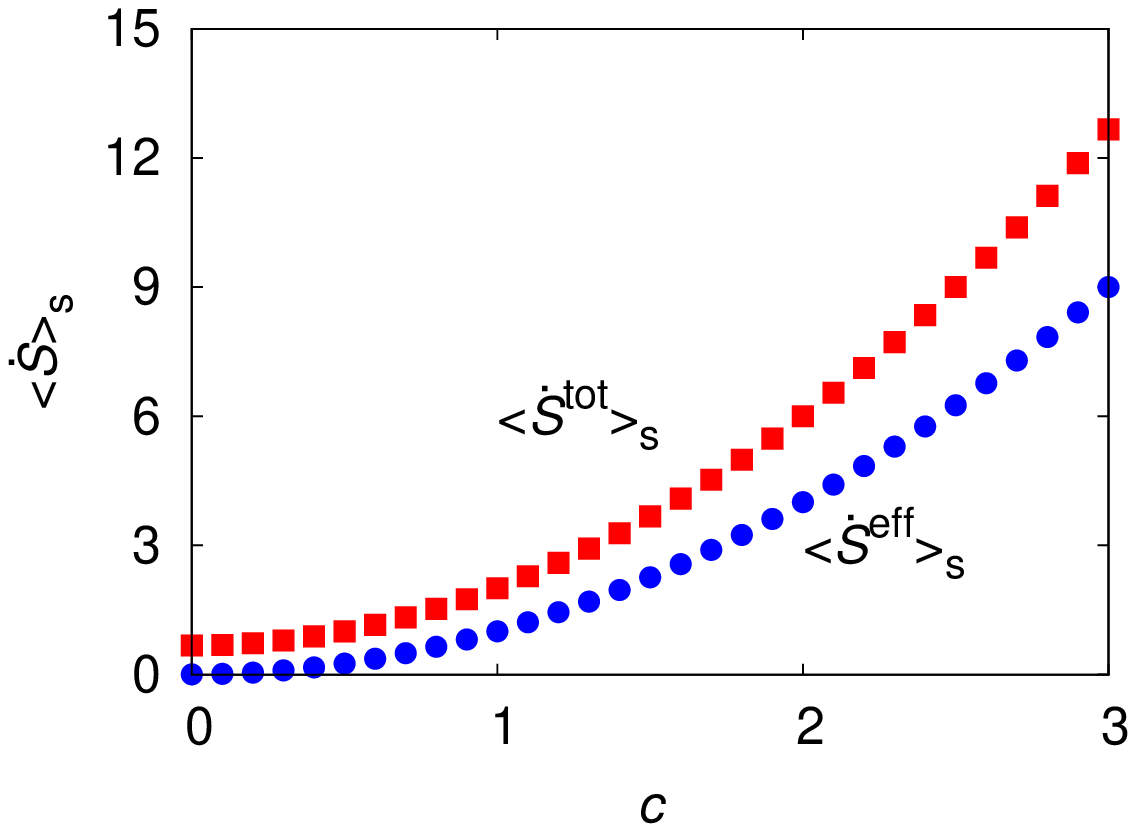}
\caption{ Total and effective entropy production rates as a function of $c$.} \label{Fig:confirm}
\end{figure}

Here, we numerically check the following second laws:
\begin{equation}
\langle dS_t^\textrm{tot} \rangle \geq \langle dS_t^\textrm{eff} \rangle \geq 0.\label{SM:inequality}
\end{equation}
As an example, consider the vane system presented in figure~\ref{transducer}(a) with $f_t(x) = -k (x-ct)$. This external force can be realized by a rotating torsional spring with stiffness $k$ and constant angular velocity $c$. By solving the Langevin equation~(1) numerically, we calculate the average rates of heat flows in the steady state as $\langle \dot{Q}^{(1)} \rangle_\textrm{s}$  and $\langle \dot{Q}^{(2)} \rangle_\textrm{s}$, where $\langle \dot{Q}^{(\nu)} \rangle_\textrm{s} \equiv \langle dQ_{t}^{(\nu)}/dt \rangle_\textrm{s}$. For the steady-state average, we collect $10^5$ data after the system reaches the steady state. From equations~\eqref{eq:1EP} and \eqref{eq:2EP}, the total and effective EP rates become
\begin{eqnarray}
\langle \dot{S}^\textrm{tot} \rangle_\textrm{s}  &= &
-\frac{\langle \dot{Q}^{(1)} \rangle_\textrm{s}}{T^{(1)}} -\frac{\langle \dot{Q}^{(2)} \rangle_\textrm{s}}{T^{(2)}}, \nonumber \\
\langle \dot{S}^\textrm{eff} \rangle_\textrm{s} &= &
-\frac{\langle \dot{Q}^{(1)} \rangle_\textrm{s} +\langle \dot{Q}^{(2)} \rangle_\textrm{s}}{T^{(1,2)}}, \label{eq:numerical_check}
\end{eqnarray}
respectively. For this calculation, we use the second-order integrator~\cite{Ciccotti} with parameters $k=1$, $\gamma^{(1)}=\gamma^{(2)}=1$, $T^{(1)}=3$, and $T^{(2)}=1$, thus $T^{(1,2)}=2$.

Figure~\ref{Fig:confirm} shows $\langle \dot{S}^\textrm{tot} \rangle_\textrm{s} $ and $\langle \dot{S}^\textrm{eff} \rangle_\textrm{s}$, denoted by squares and circles respectively, as a function of $c$. At $c=0$, we can analytically calculate the heat flux rate as $\langle \dot{Q}^{(1)} \rangle_\textrm{s}= \gamma^{(1)}\gamma^{(2)} ( T^{(1)} - T^{(2)})/(\gamma^{(1)} + \gamma^{(2)}) = -\langle \dot{Q}^{(2)} \rangle_\textrm{s}$~\cite{Murashita}. Thus, we get $\langle \dot{Q}^{(1)} \rangle_\textrm{s}= 1 = -\langle \dot{Q}^{(2)} \rangle_\textrm{s}$ along with $\langle \dot{S}^\textrm{tot} \rangle_\textrm{s} =2/3 $ and $\langle \dot{S}^\textrm{eff} \rangle_\textrm{s}=0$, which is confirmed in figure~\ref{Fig:confirm}. As the external driving becomes stronger with increasing $c$,
both $\langle \dot{S}^\textrm{tot} \rangle_\textrm{s}$ and $\langle \dot{S}^\textrm{eff} \rangle_\textrm{s}$ increase, still
satisfying the inequality~\eqref{SM:inequality}. Note that the difference between $\langle \dot{S}^\textrm{tot} \rangle_\textrm{s}$ and $\langle \dot{S}^\textrm{eff} \rangle_\textrm{s}$ is the mixing EP rate which is always larger than zero. This numerical result clearly supports our theory for the second laws of the total, effective, and mixing EPs.

\section{Hierarchy of entropy productions for a $n$-reservoir system}

\begin{figure*}
\centering
\includegraphics[width=0.7\linewidth]{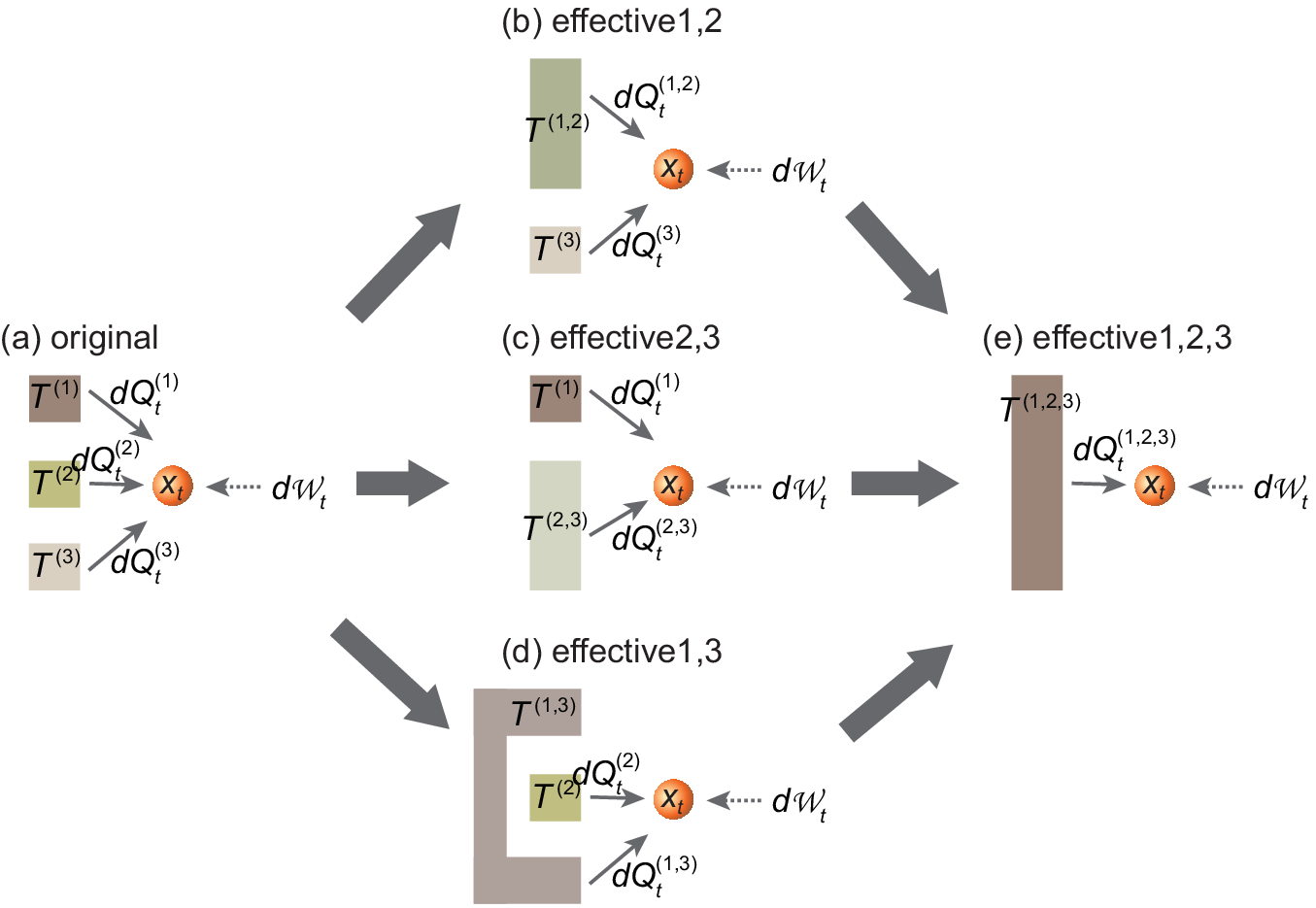}
\caption{ Hierarchy of EPs. (a) Schematic of the original system, where a system is in contact with three reservoirs simultaneously. (b), (c), and (d) are the schematics of the effective systems made by combining the two of the three reservoirs. (e) is the schematic of the effective system made by combining all  three reservoirs. } \label{Fig:hierarchy}
\end{figure*}

To understand the hierarchy of EPs for a $n$-reservoir system, consider the example illustrated in figure~\ref{Fig:hierarchy}, where the vane system is connected to $3$ reservoirs simultaneously. The EP of the original system, figure~\ref{Fig:hierarchy}(a), is given by
\begin{eqnarray}
\langle dS_t^\textrm{tot} \rangle_\textrm{s}  &= &
-\frac{\langle dQ_t^{(1)} \rangle_\textrm{s}}{T^{(1)}} -\frac{\langle dQ_t^{(2)}\rangle_\textrm{s}}{T^{(2)}}
-\frac{\langle dQ_t^{(3)}\rangle_\textrm{s}}{T^{(3)}}\geq 0. \label{eq:original_EP}
\end{eqnarray}
We can write an effective EP by combining  the reservoirs $1$ and $2$, $2$ and $3$, and $1$ and $3$ as one effective reservoir with effective temperatures $T^{(1,2)}$, $T^{(2,3)}$, and $T^{(1,3)}$ from the original system as illustrated in figures~\ref{Fig:hierarchy}(b), \ref{Fig:hierarchy}(c), and \ref{Fig:hierarchy}(d), whose EPs are given by
\begin{eqnarray}
\langle dS_t^\textrm{eff1,2} \rangle_\textrm{s}  &= &
-\frac{\langle dQ_t^{(3)}\rangle_\textrm{s}}{T^{(3)}}
-\frac{\langle dQ_t^{(1,2)} \rangle_\textrm{s}}{T^{(1,2)}} \geq 0
, \nonumber \\
\langle dS_t^\textrm{eff2,3} \rangle_\textrm{s}  &= &
-\frac{\langle dQ_t^{(1)}\rangle_\textrm{s}}{T^{(1)}}
-\frac{\langle dQ_t^{(2,3)} \rangle_\textrm{s}}{T^{(2,3)}} \geq 0 , \nonumber \\
\langle dS_t^\textrm{eff1,3} \rangle_\textrm{s}  &= &
-\frac{\langle dQ_t^{(2)}\rangle_\textrm{s}}{T^{(2)}}
-\frac{\langle dQ_t^{(1,3)} \rangle_\textrm{s}}{T^{(1,3)}} \geq 0,
\end{eqnarray}
respectively, where $dQ_t^{(i,j)} = dQ_t^{(i)} + dQ_t^{(j)}$.
Furthermore, if we combine all the three reservoirs as presented in figure~\ref{Fig:hierarchy}(e), the EP can be written as
\begin{eqnarray}
\langle dS_t^\textrm{eff1,2,3} \rangle_\textrm{s}  &= &
-\frac{\langle dQ_t^{(1,2,3)} \rangle_\textrm{s}}{T^{(1,2,3)}} \geq 0,
\end{eqnarray}
where $dQ_t^{(1,2,3)} = \sum_{i=1}^3 dQ_t^{(i)}$. Then, one can find the hierarchy of EPs as
\begin{eqnarray}
\langle dS_t^\textrm{tot} \rangle_\textrm{s} \geq \langle dS_t^\textrm{eff1,2} \rangle_\textrm{s} \geq \langle dS_t^\textrm{eff1,2,3} \rangle \geq 0, \nonumber \\
\langle dS_t^\textrm{tot} \rangle_\textrm{s} \geq \langle dS_t^\textrm{eff2,3} \rangle_\textrm{s} \geq \langle dS_t^\textrm{eff1,2,3} \rangle \geq 0, \nonumber \\
\langle dS_t^\textrm{tot} \rangle_\textrm{s} \geq \langle dS_t^\textrm{eff1,3} \rangle_\textrm{s} \geq \langle dS_t^\textrm{eff1,2,3} \rangle \geq 0. \label{eq:mixing}
\end{eqnarray}
Note that magnitude relations between $\langle dS_t^{\textrm{eff} i,j} \rangle_\textrm{s}$ are not determined by Eq.~\eqref{eq:mixing}. This procedure can be extended to a general $n$-reservoir system.

\vfil\eject


\begin{thebibliography}{99}

%De Domenico M and Arenas A 2017 Modeling structure and resilience of the dark network Phys. Rev. E 95 022313


\bibitem{Verley} Verley G and Lacoste D 2012 Fluctuation theorems and inequalities generalizing the second law of thermodynamics out of equilibrium \emph{Phys. Rev.} E \textbf{86} 051127
\bibitem{Esposito} Esposito M and Van den Broeck C 2010 Three Detailed Fluctuation Theorems \emph{Phys. Rev. Lett.} \textbf{104} 090601
\bibitem{Garcia} Garc\'{i}a-Garc\'{i}a R,  Lecomte V, Kolton A B and  Dom\'{i}nguez D 2012 Joint probability distributions and fluctuation theorems \emph{J. Stat. Mech.} P02009
\bibitem{Evans} Evans D J, Cohen E G D and Morriss G P 1993 Probability of second law violations in shearing steady states \emph{Phys. Rev. Lett.} \textbf{71} 2401
\bibitem{Gallavotti} Gallavotti G and  Cohen E G D 1995 Dynamical Ensembles in Nonequilibrium Statistical Mechanics \emph{Phys. Rev. Lett.} \textbf{74} 2694
\bibitem{Jarzynski} Jarzynski C 1997 Nonequilibrium Equality for Free Energy Differences \emph{Phys. Rev. Lett.} \textbf{78} 2690
\bibitem{Kurchan} Kurchan J 1998 Fluctuation theorem for stochastic dynamics \emph{J. Phys.} A: \emph{Math. Gen.} \textbf{31} 3719
\bibitem{Lebowitz} Lebowitz J L  and  Spohn H 1999 A Gallavotti-Cohen-Type Symmetry in the Large Deviation Functional for Stochastic Dynamics \emph{J. Stat. Phys.} \textbf{95} 333-65
\bibitem{Crooks} Crooks G E 1999  Entropy production fluctuation theorem and the nonequilibrium work relation for free energy differences \emph{Phys. Rev.} E \textbf{60} 2721


\bibitem{Blickle} Blickle V and Bechinger C 2012 Realization of a micrometre-sized stochastic heat engine \emph{Nature Phys.} \textbf{8} 143
\bibitem{Martinez} Mart\'{i}nez I A,  Rold\'{a}n \'{E}, Dinis L, Petrov D,  Parrondo J M R and Rica R A 2016 Brownian Carnot engine \emph{Nature Phys.} \textbf{12} 67–70
\bibitem{Su} Quinto-Su P A 2014 A microscopic steam engine implemented in an optical tweezer \emph{Nat. Commun.} \textbf{5} 5889

\bibitem{bacteria} Krishnamurthy S,  Ghosh S, Chatterji D, Ganapathy R and Sood A K 2016  A micrometre-sized heat engine operating between bacterial reservoirs \emph{Nature Phys.} \textbf{12} 1134–38

\bibitem{Johannes} Ro{\ss}nagel J, Dawkins S T, Tolazzi K N, Abah O, Lutz E,  Schmidt-Kaler F and Singer K 2016 A single-atom heat engine
\emph{Science} \textbf{352} 325-29

\bibitem{Schmidt} Schmidt F,  Magazz\`{u} A, Callegari A, Biancofiore L, Cichos F and Volpe G 2018 Microscopic Engine Powered by Critical Demixing \emph{Phys. Rev. Lett.} \textbf{120} 068004


\bibitem{Broeck} Van den Broeck C, Kawai R and Meurs P 2004 Microscopic Analysis of a Thermal Brownian Motor \emph{Phys. Rev. Lett.} \textbf{93} 090601
\bibitem{Kawai} Van den Broeck C and Kawai R 2006 Brownian Refrigerator \emph{Phys. Rev. Lett.} \textbf{96} 210601

\bibitem{Horowitz} Horowitz J M 2015 Diffusion approximations to the chemical master equation only have a consistent stochastic thermodynamics at chemical equilibrium \emph{J. Chem. Phys.} \textbf{143} 044111

\bibitem{Broeck1} Van den Broeck C and Esposito M 2010 Three faces of the second law. II. Fokker-Planck formulation \emph{Phys. Rev.} E \textbf{82} 011144
\bibitem{Murashita} Murashita Y and Esposito M 2016 Overdamped stochastic thermodynamics with multiple reservoirs \emph{Phys. Rev.} E \textbf{94} 062148

\bibitem{Seifert2005} Seifert U 2005 Entropy Production along a Stochastic Trajectory and an Integral Fluctuation Theorem \emph{Phys. Rev. Lett.} \textbf{95} 040602

\bibitem{Schnakenberg} Schnakenberg J 1976 Network theory of microscopic and macroscopic behavior of master equation systems \emph{Rev. Mod. Phys.} \textbf{48} 571 (1976).

\bibitem{Hatano} Hatano T and Sasa S-I 2001 Steady-State Thermodynamics of Langevin Systems \emph{Phys. Rev. Lett.} \textbf{86} 3463
\bibitem{Speck_Seifert} Speck T and Seifert U 2005 Integral fluctuation theorem for the housekeeping heat \emph{J. Phys.} A \textbf{38} L581
\bibitem{Spinney} Spinney R E  and  Ford I J 2012 Nonequilibrium Thermodynamics of Stochastic Systems with Odd and Even Variables \emph{Phys. Rev. Lett.} \textbf{108} 170603
\bibitem{Spinney1} Spinney R E  and  Ford I J 2012 Entropy production in full phase space for continuous stochastic dynamics \emph{Phys. Rev.} E \textbf{85} 051113
\bibitem{P1} Kwon C, Yeo J, Lee H K and Park H 2016 Unconventional entropy production in the presence of momentum-dependent forces
    \emph{J. Korean Phys. Soc.} \textbf{68} 633
\bibitem{P2} Yeo J, Kwon C, Lee H K and Park H 2016 Housekeeping entropy in continuous stochastic dynamics with odd-parity variable \emph{J. Stat. Mech.} 093205
\bibitem{Tome} Tom\'{e} T  and De Oliveira M J 2012 Entropy Production in Nonequilibrium Systems at Stationary States \emph{Phys. Rev. Lett.} \textbf{108} 020601
\bibitem{Sagawa} Sagawa T and Ueda M 2012 Fluctuation Theorem with Information Exchange: Role of Correlations in Stochastic Thermodynamics \emph{Phys. Rev. Lett.} \textbf{109} 180602
\bibitem{Lee} Lee H K, Kwon C and Park H 2013 Fluctuation Theorems and Entropy Production with Odd-Parity Variables \emph{Phys. Rev. Lett.} \textbf{110} 050602

\bibitem{Smoluchowski} Von Smoluchowski M 1912 Experimentell nachweisbare, der \"{u}blichen Thermodynamik widersprechende Molekularph\"{a}nomene \emph{Phys. Zeitschr.} \textbf{13} 1069-80
\bibitem{Feynman} Feynman R P 1963 \emph{The Feynman Lectures on Physics, Vol. 1.} Ch. 46, (Massachusetts, USA: Addison-Wesley)

\bibitem{MD} Lee J S and Park H 2017 Additivity of multiple heat reservoirs in Langevin equation arXiv:1712.00972
\bibitem{Visco} Visco P 2006 Work fluctuations for a Brownian particle between two thermostats \emph{J. Stat. Mech.} P06006


\bibitem{Derrida} Derrida B and  Brunet \'{E} 2005 \emph{Einstein aujourd\rq{}hui} (Les Ulis: EDP Sciences)
\bibitem{Parrondo} Parrondo J M R, Espa\~{n}ol P 1996 Criticism of Feynman’s analysis of the ratchet as an engine \emph{Am. J. Phys.} \textbf{64} 1125

\bibitem{Onsager-Machlup} Onsager L and Machlup S 1953 Fluctuations and Irreversible Processes \emph{Phys. Rev.} \textbf{91} 1505

\bibitem{Seifert_review} Seifert U 2012 Stochastic thermodynamics, fluctuation theorems and molecular machines \emph{Rep. Prog. Phys.} \textbf{75} 126001

\bibitem{Sekimoto} Sekimoto K 1998 Langevin Equation and Thermodynamics \emph{Prog. Theor. Phys.} \textbf{130} 17
\bibitem{Noh} Noh J D and Park J-M 2012 Fluctuation Relation for Heat \emph{Phys. Rev. Lett.} \textbf{108} 240603
\bibitem{jacob} The Jacobian has a correction of the order of $(dt)^2$ from the unity, which is irrelevant in the derivation of equation
\eqref{eq:FT_tot}.
\bibitem{Fogedby} Fogedby H C and Imparato A 2014 Heat fluctuations and fluctuation theorems in the case of multiple reservoirs \emph{J. Stat. Mech.} P11011

\bibitem{Broeck2} Esposito M and Van den Broeck C 2010 Three faces of the second law. I. Master equation formulation \emph{Phys. Rev.} E \textbf{82} 011143
\bibitem{Broeck3} Esposito M and Van den Broeck C 2010 Three Detailed Fluctuation Theorems \emph{Phys. Rev. Lett.} \textbf{104} 090601
\bibitem{log_sum} Cover T M and  Thomas J A 1991 {\em Elements of Information Theory} (Wiley)
\bibitem{exp} Engines with multi degrees of freedom (not all simultaneoulsy) connected to different temperature reservoirs
can do work as a normal heat engine. Furthermore, a spatially asymmetric engine (still single degree of freedom)
considered in Ref.~\cite{Broeck} can do work even though it is simultaneoulsy connected to different reservoirs. This is
possible due to its spatial structure, which cannot be described by the simple Langevin equation~\eqref{eq:Langevin_eff}.
\bibitem{LeeJS1} Note that this setup is different from that in Ref.~\cite{LeeJS},
where no noise-mixing occurs.
\bibitem{LeeJS} Lee J S  and Park H 2017 Carnot efficiency is reachable in an irreversible process \emph{Sci. Rep.} \textbf{7} 10725

\bibitem{cooling} Andresen B, Salamon P and Berry R S 1984 Thermodynamics in finite time \emph{Phys. Today} \textbf{37} No.9, 62
\bibitem{cooling1} Levy A and Kosloff R 2012 Quantum Absorption Refrigerator \emph{Phys. Rev. Lett.} \textbf{108} 070604
\bibitem{cooling2} Mari A and Eisert J 2012 Cooling by Heating: Very Hot Thermal Light Can Significantly Cool Quantum Systems \emph{Phys. Rev. Lett.}
\textbf{108} 120602
\bibitem{Ciccotti} Vanden-Eijnden E and  Ciccotti G 2006 Second-order integrators for Langevin equations with holonomic constraints \emph{Chem. Phys. Lett.} \textbf{429} 310



\end{thebibliography}
\end{document}